\documentclass[useAMS,usenatbib,twocolumn]{mn2e}
\usepackage{amsfonts}
\usepackage{amsmath}
\usepackage{epsfig}
\usepackage{wrapfig}
\usepackage{amssymb}
\usepackage{longtable}
\usepackage{accents}
\usepackage{enumitem} 
\SetLabelAlign{CenterWithParen}{\hfil({#1})\hfil}  
\usepackage{dcolumn}
\usepackage{multirow}
\usepackage{bm}
\usepackage[caption=false]{subfig}
\usepackage{hyperref}
\hypersetup{colorlinks=true,linkcolor=blue,citecolor=blue,filecolor=blue,urlcolor=blue}

\begin{document}

\title[Jet of the quasar S5~1928+738]{Signatures of a spinning supermassive black hole binary on the mas-scale jet of the quasar S5~1928+738 based on 25 years of VLBI data}
\author[E. Kun et al.]
{\parbox{\textwidth}{E. Kun$^{1,2,3,4,5}$\thanks{%
E-mail: kun.emma@csfk.org}, S. Britzen$^{6}$, S. Frey$^{4,5,7}$, K. \'{E}. Gab\'{a}nyi$^{4,5,8,9}$, L. \'{A}. Gergely$^{10,11,12}$}\\
\vspace{0.4cm}\\
$^{1}$ Theoretical Physics IV, Faculty for Physics \& Astronomy, Ruhr University Bochum, 44780 Bochum, Germany\\
$^{2}$ Faculty for Physics \& Astronomy, Astronomical Institute, Ruhr University Bochum, 44780 Bochum, Germany\\
$^{3}$ Ruhr Astroparticle And Plasma Physics Center, Ruhr-Universit\"at Bochum, 44780 Bochum, Germany\\
$^{4}$ Konkoly Observatory, HUN-REN Research Centre for Astronomy and Earth Sciences, Konkoly Thege Mikl\'os \'ut 15-17, H-1121 Budapest, Hungary\\
$^{5}$  CSFK, MTA Centre of Excellence, Konkoly Thege Miklós \'ut 15-17, H-1121 Budapest, Hungary\\
$^{6}$ Max-Planck-Institute f\"{u}r Radioastronomie, Auf dem H\"{u}gel 69, D-53121 Bonn, Germany\\
$^{7}$ Institute of Physics and Astronomy, ELTE E\"{o}tv\"{o}s Lor\'{a}nd University, P\'{a}zm\'{a}ny P\'{e}ter s\'{e}t\'{a}ny 1/A, H-1117 Budapest, Hungary\\
$^{8}$  Department of Astronomy, Institute of Physics and Astronomy, ELTE E\"{o}tv\"{o}s Lor\'{a}nd University, P\'{a}zm\'{a}ny P\'{e}ter s\'{e}t\'{a}ny 1/A, H-1117 Budapest, Hungary\\
$^{9}$  HUN-REN--ELTE Extragalactic Astrophysics Research Group, P\'{a}zm\'{a}ny P\'{e}ter s\'{e}t\'{a}ny 1/A, H-1117 Budapest, Hungary\\ 
$^{10}$Department of Experimental Physics, University of Szeged, D\'om t\'er 9, H-6720 Szeged, Hungary\\
$^{11}$Department of Theoretical Physics, University of Szeged, Tisza Lajos krt. 84-86, H-6720 Szeged, Hungary\\
$^{12}$Department of Theoretical Physics, HUN-REN Wigner Research Centre for Physics, Konkoly-Thege Mikl{\'o}s \'ut 29-33, H-1121 Budapest, Hungary}
\date{Accepted . Received ; in original form }
\maketitle

\begin{abstract}
In a previous work, we have identified the spin of the dominant black hole of a binary from its jet properties. Analysing Very Long Baseline Array (VLBA) observations of the quasar S5~1928+738, taken at $15$-GHz during $43$ epochs between $1995.96$ and $2013.06$, we showed that the inclination angle variation of the inner ($<2$~mas) jet symmetry axis naturally decomposes into a periodic and a monotonic contribution. The former emerges due to the Keplerian orbital evolution, while the latter is interpreted as the signature of the spin-orbit precession of the jet emitting black hole.  
In this paper, we revisit the analysis of the quasar S5~1928+738 by including new $15$-GHz VLBA observations extending over $29$ additional epochs, between $2013.34$ and $2020.89$. The extended data set confirms our previous findings which are further supported by the flux density variation of the jet. By applying an enhanced jet precession model that can handle arbitrary spin orientations $\kappa$ with respect to the orbital angular momentum of a binary supermassive black hole system, we estimate the binary mass ratio as $\nu=0.21\pm0.04$ for $\kappa=0$ (i.e. when the spin direction is perpendicular to the orbital plane) and as $\nu=0.32\pm0.07$ for $\kappa=\pi/2$ (i.e. when the spin lies in the orbital plane). 
We estimate more precisely the spin precession velocity, halving its uncertainty from $(-0.05\pm0.02)\degr\,\mathrm{yr}^{-1}$ to $(-0.04\pm0.01)\degr\,\mathrm{yr}^{-1}$.
\end{abstract}

\pagerange{\pageref{firstpage}--\pageref{lastpage}}

\label{firstpage}


\section{Introduction}

When supermassive black holes (SMBHs), located in the centre of large galaxies merge as their host galaxies undergo a major merger \citep[e.g.][]{Begelman1980}, first the dynamical friction, then the gravitational radiation governs the dynamical evolution of their orbit \citep[e.g.][]{BinneyTremaine1987, MerrittMilos2005}. Once the dissipation of energy and angular momentum becomes dominated by gravitational radiation, the already sub-pc separated binary goes through three main phases: inspiral, plunge and ringdown \citep[e.g.][]{Kidder1995}. Post-Newtonian (PN) techniques are involved in the description of the binary dynamics in the inspiral phase for which the PN parameter is constrained as $0.001<\varepsilon<0.1$ \citep[e.g.][]{Gergely2009}. In the plunge stage, numerical calculations are required to characterise the final SMBH \citep[e.g.][]{Buonanno2000}. The ringdown is characterised in terms of quasinormal modes of black holes \citep{Berti2009}.

A large number of archival studies identified indirect signals of sub-pc  separated SMBH binaries in radio-loud active galactic nuclei (AGN), including the recent works \citep[e.g.][]{Eracleous2012,Graham2015,Liu2016,Charisi2016,Chen2020}. SMBH binaries are also emerging in the multimessenger discussion of AGN, mostly of blazars \citep[e.g.][]{Kun2019,Britzen2019,deBruijn2020,Britzen2021,jaroschewski2022,BeckerTjus2022,Britzen2023}.

The source we study here, S5~1928+738 \citep[redshift $z_\mathrm{spect}=0.302$, ][]{Lawrence1986} is a low-synchrotron-peaked, low polarization quasar ($\mathrm{RA}=19^\mathrm{h} 27^\mathrm{m} 48\fs495$, $\mathrm{Dec}=+73\degr 58\arcmin 1\farcs570$), with an elongated jet structure, extending over $\sim 20$ mas at $15$-GHz. Three decades ago, the presence of a SMBH binary was already suspected at the centre of this quasar \citep{Roos1993}. Based on the long-term data  of the jet of the quasar S5~1928+738 from the Monitoring Of Jets in Active galactic nuclei with VLBA Experiments (MOJAVE) program, obtained with the Very Long Baseline Array (VLBA), we further strengthened the idea that this quasar harbours a SMBH binary \citep{Kun2014}. In that work, we have shown the spin-precession of the dominant black hole. We monitored the direction of the symmetry axis of the inner $2$ milliarcseconds (mas) of the jet of S5~1928+738 through its inclination and position angles. By fitting the shape of a projected helical jet to the position of the VLBA-detected jet components, we derived the value of the inclination and position angles. Then we decomposed the observed time variation of the inclination of the symmetry axis into a periodic term with an amplitude of $\sim 0\fdg89$ and a linear decreasing trend with a rate of $\sim 0.05\degr\,\mathrm{yr}^{-1}$.

We explained the periodic component in the direction of the inner jet as generated by the SMBH binary with total mass $m\approx{8.13 \times 10^{8}} M_{\odot}$ ($m=m_1+m_2$, $m_2<m_1$). This periodic change in the direction of the symmetry axis of the jet is induced by the orbital motion of the jet emitter dominant black hole. We derived the binary parameters consistent with the VLBA measurements spanning through $18$ years: orbital period $T=(4.78\pm0.14)$~yr, separation $r=(0.0128 \pm 0.0003)$~pc and PN parameter $\varepsilon\approx 0.003$, lower limit of the mass ratio {$\nu \gtrsim 1/5$} ($\nu=m_2/m_1$). The upper limit of $\nu$ could not be estimated from the observational data. Therefore, we adopted the upper limit of the mass ratios typical for SMBH mergers, $\nu \lesssim 1/3$ \citep{Gergely2009}. 

Next, we identified the slow, secular reorientation of the jet (superposed to the periodic component) as due to the spin-orbit precession of the jet emitter SMBH. From the observations, we derived an upper limit for the precession period of the more massive black hole ($T_{\mathrm{SO}}< 5500$~yr) and the gravitational time scale of the merger ($T_{\mathrm{GW}}<1.64 \times 10^6$~yr). These results apply for zero spin angle ($\kappa=0$), i.e., assuming the spin of the jet emitter SMBH perpendicular to the orbital plane.

Periodic jet structures are attributed either to the orbital motion of supermassive black hole binaries \citep{Hummel1992,Roos1993,Britzen2001,Deane2014}, or instabilities in the plasma. Our model presented in \citet{Kun2014} introduces a scenario in which the orbital motion of the jet emitter black hole shakes up an intrinsically helical jet \citep[for visualisation, see Figure 9. in][]{Kun2018}.

In this paper, we extend the time range of the $15$-GHz very long baseline interferometry (VLBI) observations of the jet of S5~1928+738, by adding new VLBA observations taken in $29$ epochs between $2013.34$ and $2020.89$. In Section~\ref{vlbadata}, we describe the analysis of these MOJAVE data. In Section~\ref{morphkin}, we investigate the overall jet morphology and kinematics between $1995.96$ and $2020.89$. We study the structure of the inner $2$~mas of the jet of S5~1928+738 to derive the variation of the inclination angle of the inner jet symmetry axis. In Section~\ref{sec:model}, we develop a new binary SMBH jet model enabling arbitrary spin angles. In Section~\ref{sec:test}, we apply this model to the inclination angle variation of the inner jet axis of S5~1928+738 and constrain the mass ratio as function of the spin angle $\kappa$, generalising our earlier results. In this section, we also present a sanity check with the 15-GHz flux densities. Finally, in Section~\ref{discsum}, we discuss and summarise our results. 
We adopt the cosmological parameters as follows: $H_0=67.8$~km\,s$^{-1}$\,Mpc$^{-1}$, $\Omega_\mathrm{m}=0.308$ and $\Omega_\Lambda=0.692$.

\begin{table*}
\begin{center}
\caption{Summary of the image parameters of the observing epochs between 2013.34 and 2019.65 for S5~1928+738. (1) Date of observation, (2) VLBA project code, (3) participating VLBA stations BR – Brewster, FD – Fort Davis, HN – Hancock, KP – Kitt Peak, LA – Los Alamos, MK – Mauna Kea, NL– North Liberty, OV – Owens Valley, PT – Pie Town, SC – St. Croix (`All' means all telescopes are participating, and e.g. '--BR' means `All but Brewster'), (4) peak brightness, (5) total flux density, (6) root mean square (rms) of the brightness in the residual image, (7) major axis of the elliptical Gaussian restoring beam (FWHM), (8) minor axis of the restoring beam (FWHM), (9) position angle of the major axis of the restoring beam, (10) number of jet components, (11) reduced $\chi^2$ of the model fit. The full table is available in electronic format online.}
\begin{tabular}{lcccccccccc}
\hline
\hline
Date$^{(1)}$ & Code$^{(2)}$ & Array$^{(3)}$ & $S_\mathrm{p}^{(4)}$ & $S_\mathrm{tot}^{(5)}$ & rms$^{(6)}$ & $b_\mathrm{maj}^{(7)}$ & $b_\mathrm{min}^{(8)}$ & $b_\mathrm{pa}^{(9)}$ & $N^{(10)}$ & ${\chi^2_\mathrm{red}}^{(11)}$ \\
 & & & (Jy beam$^{-1}$) & (Jy) & (mJy beam$^{-1}$) & (mas) & (mas) & ($\degr$) & & \\
 \hline
2013-05-05 & BL178BC & All & $2.115$ & $3.405$ & $0.17$ & $0.751$ & $0.715$ & $-83.4$ & $12$ & $1.09$\\
2013-08-12 & BL178BH & $-$FD & $2.300$ & $3.626$ & $0.20$ & $0.647$ & $0.588$ & $29.0$ & $11$ & $1.07$\\
2013-12-15 & BL193AA & All & $2.848$ & $4.086$ & $0.09$ & $0.835$ & $0.704$ & $43.4$ & $11$ & $1.27$\\
2014-09-01 & BL193AN & All & $3.675$ & $4.878$ & $0.12$ & $0.653$ & $0.621$ & $-35.1$ & $12$ & $1.36$\\
2015-06-16 & BL193AT & All & $5.199$ & $6.350$ & $0.10$ & $0.702$ & $0.628$ & $-22.5$ & $10$ & $1.92$\\
 \hline
\hline
\label{vlbatable}
\end{tabular}
\end{center}
\end{table*}

\begin{table*}
\caption{Circular Gaussian model fit results for S5~1928+738. (1) epoch of observation, (2) flux density, (3)--(4) relative right ascension and declination of the component centre with respect to the core, (5) component diameter (FWHM), (6) jet component identifier. The full table is available in electronic format online.}
\begin{tabular}{lccccc}
\hline
\hline
Epoch$^{(1)}$ & $S^{(2)}$ & $x^{(3)}$ & $y^{(4)}$  & $d^{(5)}$  & ID$^{(6)}$\\ 
(yr) & (Jy) & (mas) & (mas) & (mas)  & \\ 
\hline
2013.34 & $ 0.241 \pm 0.009 $ & $ 0.000 \pm 0.019 $ & $ 0.000 \pm 0.018 $ & $ 0.023 \pm 0.009 $ & CS\\
 & $ 1.855 \pm 0.025 $ & $ 0.080 \pm 0.020 $ & $ -0.240 \pm 0.038 $ & $ 0.128 \pm 0.001 $ & Cg\\
 & $ 0.505 \pm 0.013 $ & $ 0.306 \pm 0.021 $ & $ -0.774 \pm 0.020 $ & $ 0.169 \pm 0.004 $ & C16\\
 & $ 0.346 \pm 0.011 $ & $ 0.363 \pm 0.021 $ & $ -1.057 \pm 0.020 $ & $ 0.182 \pm 0.006 $ & C15\\
 & $ 0.145 \pm 0.007 $ & $ 0.646 \pm 0.024 $ & $ -1.709 \pm 0.023 $ & $ 0.288 \pm 0.017 $ & C14\\
 & $ 0.054 \pm 0.005 $ & $ 1.074 \pm 0.039 $ & $ -2.217 \pm 0.039 $ & $ 0.694 \pm 0.069 $ & C13\\
 & $ 0.100 \pm 0.006 $ & $ 1.584 \pm 0.032 $ & $ -3.149 \pm 0.032 $ & $ 0.529 \pm 0.030 $ & C11\\
 & $ 0.080 \pm 0.006 $ & $ 1.375 \pm 0.051 $ & $ -3.800 \pm 0.050 $ & $ 0.938 \pm 0.066 $ & C9\\
 & $ 0.037 \pm 0.004 $ & $ 1.517 \pm 0.059 $ & $ -4.946 \pm 0.059 $ & $ 1.127 \pm 0.187 $ & C8\\
 & $ 0.013 \pm 0.003 $ & $ 1.442 \pm 0.065 $ & $ -6.747 \pm 0.065 $ & $ 1.240 \pm 0.326 $ & C7\\
 & $ 0.013 \pm 0.003 $ & $ 2.469 \pm 0.082 $ & $ -9.089 \pm 0.082 $ & $ 1.595 \pm 0.855 $ & C4\\
 & $ 0.016 \pm 0.006 $ & $ 2.393 \pm 0.141 $ & $ -17.060 \pm 0.141 $ & $ 2.789 \pm 2.973 $ & C1\\
\hline
\hline
\label{comptable}
\end{tabular}
\end{table*}

\begin{figure*}
\begin{center}
\includegraphics[scale=0.5,angle=270]{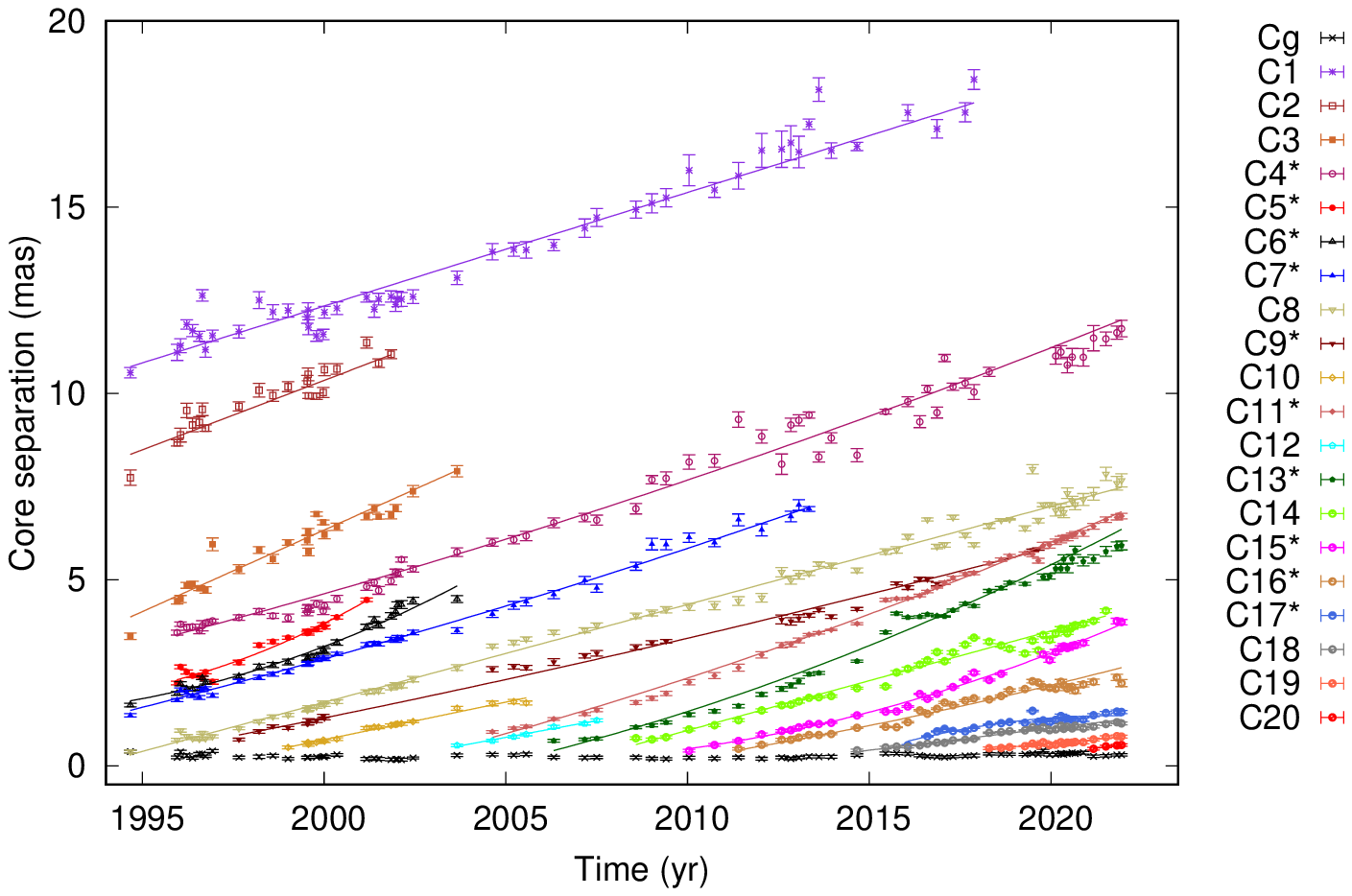}
\end{center}
\caption{Core separation of the jet components as a function of time with error bars and their fitted proper motion represented by continuous lines. Accelerating components are marked by $\star$. The separation of the jet components is relative to the VLBI core `CS' at $(0,0)$.}
\label{coresep}
\end{figure*}
\begin{table*}
\caption{Component speed fits. (1) jet component identifier. From linear fit (L): (2) linear proper motion, (3) ejection time, (4) $\chi^2$. From accelerated fit (A): (5) non-linear proper motion, (6) acceleration rate, (7) ejection time, (8) $\chi^2$. (9) Difference between the AICs ($\Delta=\mathrm{AIC}_\mathrm{lin}-\mathrm{AIC}_\mathrm{acc}$), (10) type of the proper motion based on the AIC (L - linear, A - accelerated), see details in the text, (11) apparent speed.}
\begin{tabular}{ccccccccccc}
\hline
\hline
ID$^{(1)}$ & $\mu_\mathrm{L}^{(2)}$ & $t_\mathrm{ej,L}^{(3)}$ & ${\chi^2}^{(4)}$& $\mu_\mathrm{A}^{(5)}$ & ${\dot{\mu}_\mathrm{A}}^{(6)}$ & $t_\mathrm{ej,A}^{(7)}$ & ${\chi^2}^{(8)}$ & ${\Delta}^{(9)}$ & PM$^{(10)}$ & ${\beta_\mathrm{app,lin}}^{(11)}$\\ 
 & (mas yr$^{-1}$) & (yr) & & (mas yr$^{-1}$) & (mas yr$^{-2}$) & (yr) & & & & ($c$) \\ 
\hline
C1 & $ 0.305 \pm 0.010 $ & $ 1959.5 \pm 1.4 $ & $ 270.2 $ & $ 0.312 \pm 0.009 $ & $ 0.011 \pm 0.003 $ & $ 1961.4 \pm 1.3 $ & $ 261.0 $ & $ 7.2 $ & L & $ 5.97 \pm 0.20 $\\
C2 & $ 0.373 \pm 0.032 $ & $ 1972.3 \pm 2.3 $ & $ 102.1 $ & $ 0.375 \pm 0.034 $ & $ -0.011 \pm 0.037 $ & $ 1972.4 \pm 2.3 $ & $ 101.6 $ & $ -1.5 $ & L & $ 7.30 \pm 0.63 $\\
C3 & $ 0.438 \pm 0.026 $ & $ 1985.5 \pm 0.8 $ & $ 192.1 $ & $ 0.407 \pm 0.024 $ & $ -0.066 \pm 0.021 $ & $ 1984.2 \pm 0.9 $ & $ 187.3 $ & $ 2.8 $ & L & $ 8.57 \pm 0.51 $\\
C4 & $ 0.315 \pm 0.005 $ & $ 1985.2 \pm 0.3 $ & $ 593.4 $ & $ 0.325 \pm 0.006 $ & $ 0.006 \pm 0.002 $ & $ 1986.4 \pm 0.5 $ & $ 512.8 $ & $ 78.6 $ & A & $ 6.16 \pm 0.10 $\\
C5 & $ 0.392 \pm 0.019 $ & $ 1990.2 \pm 0.4 $ & $ 146.2 $ & $ 0.407 \pm 0.020 $ & $ 0.058 \pm 0.030 $ & $ 1990.8 \pm 0.4 $ & $ 119.6 $ & $ 24.6 $ & A & $ 7.67 \pm 0.37 $\\
C6 & $ 0.328 \pm 0.013 $ & $ 1990.0 \pm 0.3 $ & $ 162.3 $ & $ 0.343 \pm 0.012 $ & $ 0.038 \pm 0.010 $ & $ 1990.7 \pm 0.3 $ & $ 104.0 $ & $ 56.3 $ & A & $ 6.42 \pm 0.25 $\\
C7 & $ 0.287 \pm 0.004 $ & $ 1989.9 \pm 0.2 $ & $ 144.1 $ & $ 0.292 \pm 0.004 $ & $ 0.005 \pm 0.001 $ & $ 1990.3 \pm 0.2 $ & $ 105.6 $ & $ 36.5 $ & A & $ 5.62 \pm 0.08 $\\
C8 & $ 0.263 \pm 0.002 $ & $ 1993.5 \pm 0.1 $ & $ 760.2 $ & $ 0.263 \pm 0.002 $ & $ -0.001 \pm 0.001 $ & $ 1993.4 \pm 0.2 $ & $ 758.1 $ & $ 0.1 $ & L & $ 5.15 \pm 0.04 $\\
C9 & $ 0.224 \pm 0.003 $ & $ 1994.3 \pm 0.2 $ & $ 328.1 $ & $ 0.225 \pm 0.003 $ & $ 0.003 \pm 0.001 $ & $ 1994.8 \pm 0.3 $ & $ 278.2 $ & $ 47.9 $ & A & $ 4.38 \pm 0.06 $\\
C10 & $ 0.202 \pm 0.006 $ & $ 1996.5 \pm 0.2 $ & $ 22.8 $ & $ 0.198 \pm 0.006 $ & $ -0.013 \pm 0.005 $ & $ 1996.3 \pm 0.2 $ & $ 16.8 $ & $ 4.0 $ & L & $ 3.96 \pm 0.12 $\\
C11 & $ 0.341 \pm 0.005 $ & $ 2002.8 \pm 0.2 $ & $ 523.9 $ & $ 0.346 \pm 0.003 $ & $ 0.010 \pm 0.002 $ & $ 2003.3 \pm 0.1 $ & $ 247.7 $ & $ 274.2 $ & A & $ 6.71 \pm 0.08 $\\
C12 & $ 0.180 \pm 0.009 $ & $ 2000.7 \pm 0.3 $ & $ 3.2 $ & $ 0.179 \pm 0.009 $ & $ 0.013 \pm 0.016 $ & $ 2000.8 \pm 0.3 $ & $ 2.7 $ & $ -1.5 $ & L & $ 3.52 \pm 0.18 $\\
C13 & $ 0.367 \pm 0.010 $ & $ 2005.9 \pm 0.3 $ & $ 1303.1 $ & $ 0.375 \pm 0.009 $ & $ 0.018 \pm 0.005 $ & $ 2006.5 \pm 0.3 $ & $ 995.3 $ & $305.8 $ & A & $ 7.18 \pm 0.18 $\\
C14 & $ 0.266 \pm 0.009 $ & $ 2006.4 \pm 0.3 $ & $ 989.3 $ & $ 0.267 \pm 0.009 $ & $ -0.003 \pm 0.005 $ & $ 2006.4 \pm 0.3 $ & $ 985.7 $ & $ 1.6 $ & L & $ 5.22 \pm 0.16 $\\
C15 & $ 0.267 \pm 0.010 $ & $ 2009.3 \pm 0.2 $ & $ 705.4 $ & $ 0.268 \pm 0.007 $ & $ 0.023 \pm 0.005 $ & $ 2009.6 \pm 0.2 $ & $ 329.9 $ & $ 373.6 $ & A & $ 5.32 \pm 0.20 $\\
C16 & $ 0.209 \pm 0.009 $ & $ 2009.8 \pm 0.3 $ & $ 619.3 $ & $ 0.208 \pm 0.009 $ & $ 0.011 \pm 0.007 $ & $ 2009.9 \pm 0.3 $ & $ 591.6 $ & $ 25.7 $ & A & $ 4.05 \pm 0.16 $\\
C17 & $ 0.140 \pm 0.017 $ & $ 2010.7 \pm 0.9 $ & $ 484.6 $ & $ 0.127 \pm 0.015 $ & $ -0.071 \pm 0.023 $ & $ 2009.5 \pm 1.1 $ & $ 350.3 $ & $ 132.3 $ & A & $ 2.60 \pm 0.27 $\\
C18 & $ 0.125 \pm 0.006 $ & $ 2011.8 \pm 0.3 $ & $ 120.5 $ & $ 0.127 \pm 0.006 $ & $ -0.009 \pm 0.008 $ & $ 2011.8 \pm 0.3 $ & $ 96.3 $ & $ 22.2 $ & L & $ 2.31 \pm 0.10 $\\
C19 & $ 0.067 \pm 0.012 $ & $ 2011.2 \pm 1.5 $ & $ 18.0 $ & $ 0.068 \pm 0.012 $ & $ -0.020 \pm 0.030 $ & $ 2011.2 \pm 1.5 $ & $ 15.8 $ & $ 0.1 $ & L & $ 1.70 \pm 0.16 $\\
C20 & $ 0.097 \pm 0.013 $ & $ 1994.1 \pm 1.1 $ & $ 0.2 $ & $ 0.096 \pm 0.006 $ & $ 0.076 \pm 0.013 $ & $ 1994.3 \pm 0.5 $ & $ 0.0 $ & $ -1.8 $ & L & $ 2.49 \pm 0.33 $\\
 \hline
\hline
\label{velocityfits}
\end{tabular}
\end{table*}

\section{VLBA data analysis}
\label{vlbadata}

We analysed calibrated visibilities in 29 epochs between 2013.34 and 2020.89 taken with VLBA at $15$~GHz observing frequency, provided by the MOJAVE program \citep{Lister2009}. All ten VLBA stations participated in the observations at $19$ epochs, while single antennas were not available in $10$ epochs (see Table~\ref{vlbatable}). MOJAVE publishes amplitude- and phase-calibrated visibility data and we performed the imaging and model fitting by employing the \textsc{Difmap} software \citep{Shepherd1997}. All non-flagged data were used. The summary of the image parameters of the observing epochs between 2013.34 and 2020.89 for S5~1928+738 is given in Table~\ref{vlbatable}.

To minimise the degrees of freedom of the model fitting, we used circular Gaussian components only to describe the surface brightness distribution of the jet. The fitted parameters were the total flux density, the position, and the full width at half maximum (FWHM) size of the components. The errors were calculated as in \citet{Kun2014} and \citet{Kun2015}. The components were identified by requiring their appearance in at least $5$ subsequent epochs and smooth changes of their fitted parameters (especially the positions and flux densities). As in \citet{Kun2014}, the jet components are labelled with letter ‘C’ and a number, such that the larger number denotes a component closer to the VLBI core. This way it was straightforward to label new jet components, that were ejected after the last observing epoch considered in our earlier work \citep{Kun2014}. The results of the model fits for the epochs between 2013.34 and 2020.89 are listed in Table~\ref{comptable}.

In its core region at $15$~GHz \citep[e.g.][]{Kun2014} and $43$~GHz observing frequencies \citep[e.g.][]{Lister2000}, S5~1928+738 includes two components `CS' and `Cg'. As in our earlier work \citep{Kun2014}, the component CS is assumed to be the VLBI core. 

\section{Morphology, kinematics and jet geometry}
\label{morphkin}

\subsection{Morphology and proper motions}

The jet structure of S5~1928+738 is very rich \citep[see Figure 1 of][]{Kun2014} and the jet itself extends up to $18$ mas projected radial distance at $15$~GHz observing frequency, measured from the VLBI core. In Fig.~\ref{coresep} we plot the core separation of the jet components as a function of time. Additionally to the two quasi-stationary components CS and Cg, we identified $20$ outward-moving components.

We calculated the proper motion of the components first by fitting linear proper motions as
\begin{equation}
r_\mathrm{L}=\mu (t-t_\mathrm{ej}),
\label{eq:betalin}
\end{equation}
and then by fitting accelerating proper motions as \citep[e.g.][]{Homan2001,Lister2009}
\begin{equation}
r_\mathrm{A}=r_\mathrm{L} + \frac{\dot{\mu}}{2} \left( t-t_\mathrm{mid}\right)^2,
\label{eq:betaapp}
\end{equation}
where $\mu$ is the linear proper motion measured in mas\,yr$^{-1}$, $t_\mathrm{ej}$ the ejection time, $\dot{\mu}$ the angular acceleration and $t_\mathrm{mid}$ the half of the sum of the maximum and the minimum epochs when the respective components are detected. The proper motion of the components can be converted to apparent speed in the units of the speed of light, such that
\begin{equation}
\beta_\mathrm{app}=0.0158 \frac{\mu D_L}{(1+z_\mathrm{spect})},
\end{equation}
where $\mu$ is the proper motion, $D_L$ the luminosity distance in Mpc and $z_\mathrm{spect}$ the redshift. We summarise the linear and accelerated proper motions in Table~\ref{velocityfits}.

In order to decide whether linear or accelerated motion gives the better fit to the proper motion of jet components, we applied the Akaike Information Criterion (AIC) method \citep[][]{Akaike1974}. AIC is useful when model comparison by exact statistical tests gives inconclusive results, or for nested models. The linear proper motion (equation \ref{eq:betalin}) is nested into the accelerated proper motion (equation \ref{eq:betaapp}). First we calculated $\mathrm{AIC}=\chi^2+2N$ (where $N$ is the number of fitted parameters in a model) for the linear ($\mathrm{AIC}_\mathrm{lin}$) and acceleration model ($\mathrm{AIC}_\mathrm{acc}$) from component to component, then compared them $(\Delta=\mathrm{AIC}_\mathrm{lin}-\mathrm{AIC}_\mathrm{acc})$. The following thresholds are usually employed in the literature \citep[see e.g. in][]{Burnham2003}: $\Delta\leq2$ indicates approximately comparable performances, 
$4\leq \Delta \leq 7$ indicates a measurable difference in the fits, while $\Delta>10$ clearly favours one fit over the other. From these definitions, we only use $\Delta>10$ to distinguish between the models.
After the calculation of $\Delta$, ten components emerged as linearly moving components (C1, 2, 3, 8, 10, 12, 14, 18, 19, 20), while ten other components (C4, 5, 6, 7, 9, 11, 13, 15, 16, 17) showed signs of acceleration.

Applying different core separation upper cuts ($r\leq2$ mas, $r\leq3$ mas, $r\leq4$ mas, $r\leq5$ mas), we recalculated the $\Delta$ values between the performances of the linear and accelerating proper motion fits for the accelerating jet components, that start their observed trajectory in the innermost jet region. We show the results in Table \ref{table:diffdelta}. It seems that, for most of the components, increasing the cut-off core separation makes the linear proper motion fits worse compared to the accelerating proper motion fits. In the core separation region $r\leq2$ mas, for four components out of six, the linear proper motion fit gives better performance ($\Delta\leq 10$). This might be a sign of an acceleration zone somewhere between $2-3$~mas, therefore we adopt the core separation cut of $2$ mas from our earlier work \citep{Kun2014} to define the inner jet in S5 1928+738, where jet precession is supposed to be mostly unaltered by external forces.

The apparent speed of the jet components is related to the bulk jet speed $\beta$ and their inclination angle $\iota$, as \citep{Urry1995}
\begin{equation}
\beta_\mathrm{app}=\frac{\beta \sin \iota}{1-\beta \cos \iota},
\end{equation}
where the jet speed $\beta$ is related to the Lorentz factor $\gamma$ as
\begin{equation}
\beta=\sqrt{1-\frac{1}{\gamma^2}}.
\end{equation}
The maximal $\beta_\mathrm{app}$ occurs for the critical inclination $\cos \iota_c=\beta$, $\sin \iota_c=\gamma^{-1}$, having the value $\beta_\mathrm{app,max}=(\gamma^2-1)^{1/2}$.

In practice, when analysing the jet components, it is unlikely that any of them will have exactly the direction $\cos \iota_c=\beta$, $\sin \iota_c=\gamma^{-1}$. Hence, the jet component with maximal speed $ \beta_\mathrm{app,max} ^\mathrm{obs}\leq\beta_\mathrm{app,max}$ leads to a lower limit 
\begin{equation}
\gamma_\mathrm{low}=\sqrt{1+(\beta_\mathrm{app,max}^\mathrm{obs})^2}
\label{eq:gammamin}
\end{equation}
of the Lorentz factor of the jet.

The highest linear proper motion for the time period between 2013.34 and 2020.89, $\mu_\mathrm{max}=(0.44\pm0.03)$~mas\,yr$^{-1}$, is exhibited by component C3. This is in good agreement with our earlier results for a time period between 1995.96 and 2013.06, when we found C3 as the fastest component with the linear proper motion $\mu_\mathrm{max}=(0.43\pm0.02)$~mas\,yr$^{-1}$. With redshift $z_\mathrm{spect}=0.302$, luminosity distance $D_L=1612.6$ Mpc, this proper motion translates to $\beta_\mathrm{app,max} ^\mathrm{obs}=8.60\pm0.51$. 
For the minimum Lorentz factor given by equation (\ref{eq:gammamin}), we obtain $\gamma_\mathrm{low}\approx8.66$ and the minimum intrinsic jet speed is $\beta_\mathrm{low}\approx0.992$ in the units of the speed of the light. Then the related critical inclination angle is $\iota_\mathrm{c}=\arcsin (\gamma_\mathrm{low}^{-1})\approx 7\degr$.

\begin{table}
\caption{Differences in the $\mathrm{AIC}_\mathrm{lin}$ and $\mathrm{AIC}_\mathrm{acc}$ for the components ($\Delta=\mathrm{AIC}_\mathrm{lin}-\mathrm{AIC}_\mathrm{acc}$), that show accelerating motion, in different regions of the jet.}
\label{table:diffdelta}
\begin{tabular}{ccccc}
\hline
\hline
ID &$\Delta_\mathrm{r\leq2mas}$&$\Delta_\mathrm{r\leq3mas}$&$\Delta_\mathrm{r\leq4mas}$&$\Delta_\mathrm{r\leq 5 mas}$\\
\hline
C9 & $-1.5$ & $-2.0$ & 29.2 & 14.2\\
C11 & $-1.1$	& 28.3 & 98.4 &239.0\\
C13 & 2.8 & 99.9 & 698.9 & 424.5\\
C15 & 2.7 & 151.4 & 373.6 & 373.6\\
C16 & 42.8 & 25.7 & 25.7 & 25.7\\
C17 & 132.3 & 132.3 & 132.3 & 132.3\\
 \hline
\hline
\end{tabular}
\end{table}

\subsection{Evolution of the orientation of the inner jet symmetry axis between 1995.96 and 2020.89}
\label{sec:jetorientation}

\begin{figure*}
    \centering
    \includegraphics[angle=270,width=0.8\textwidth]{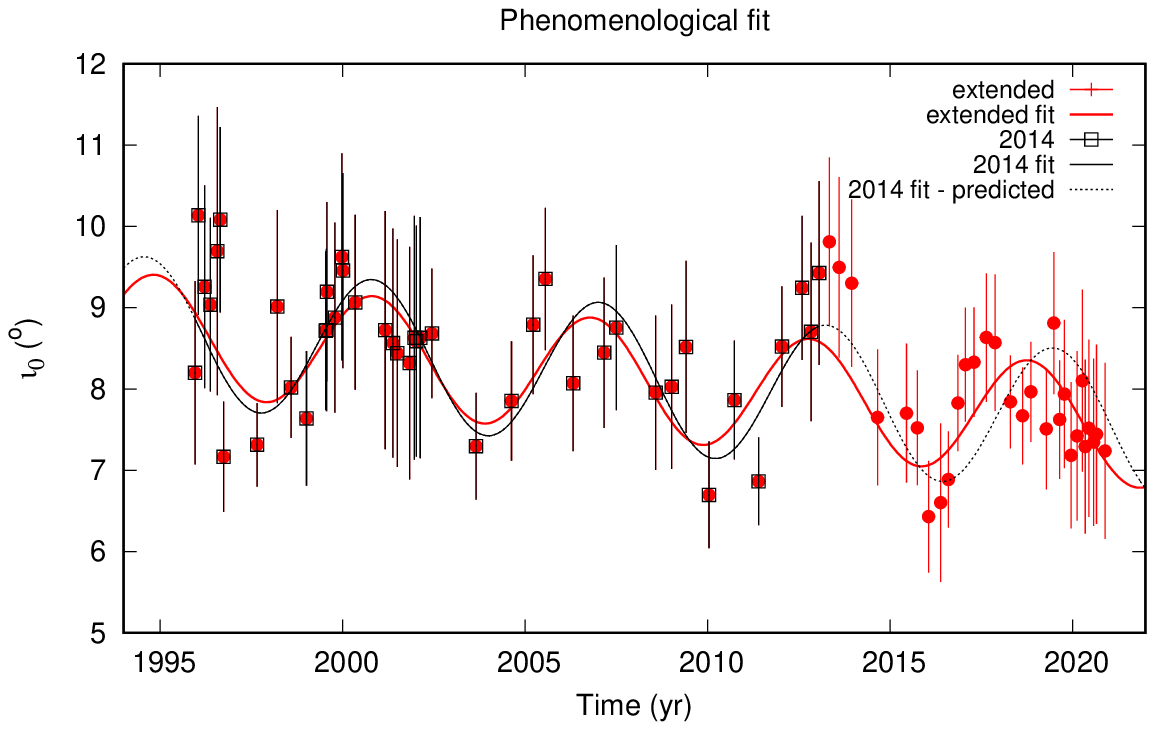}
    \caption{Phenomenological fit of the inclination angle variation of the inner jet symmetry axis of S5~1928+738 at $15$~GHz. Data points already published in \citet{Kun2014} (black squares with error bars), the phenomenological fit to them (black continuous line, ${\chi^2}_\mathrm{red}=0.55$, $D=38$, where D is the degrees of freedom), the prediction based on it (black dotted line), the new data points (red filled circles with error bars) and the phenomenological fit to the whole dataset (red continuous line, ${\chi^2}_\mathrm{red}=0.56$, D=67) are shown.}
    \label{fig:phenofit}
\end{figure*}

\begin{table*}
\caption{The parameters of the analytical fits to the time series of the inclination angles from \citet{Kun2014} and for the extended dataset. These parameters are the average ($A_0$), the amplitude of the variation ($A_1$), monotonic term ($A_2$), period of the variation ($T$) and initial phase ($\phi$), everything in the framework of the observer.}
\begin{tabular}{ccccc}
\hline
\hline
\multicolumn{5}{c}{$\iota_0(t)$, 1995.96--2013.06 \citep[][]{Kun2014}, ${\chi^2}_\mathrm{red}=0.55$}\\
\hline
$A_0[\degr]$  & $A_1[\degr]$  & $A_2$ [$\degr$\,yr$^{-1}$] & $T$ [yr] & $\phi$ [$\degr$]\\
8.67 $\pm$ 0.19 & 0.89 $\pm$ 0.17 & -0.05 $\pm$ 0.02 & 6.22 $\pm$ 0.19 & 11.95 $\pm$ 15.01\\
\hline
\multicolumn{5}{c}{$\iota_0(t)$, 1995.96--2020.89, extended, ${\chi^2}_\mathrm{red}=0.56$}\\
\hline
$A_0[\degr]$  & $A_1[\degr]$  & $A_2$ [$\degr$\,yr$^{-1}$] & $T$ [yr] & $\phi$ [$\degr$]\\
 8.64  $\pm$ 0.16 & 0.72 $\pm$ 0.11 & -0.04 $\pm$ 0.01 & 5.98$\pm$ 0.10 & 24.63$\pm$ 16.1\\
\hline
\hline
\label{fit_res}
\end{tabular}
\end{table*}
 
In \citet{Kun2014}, we described the shape of the inner jet (core separation $<2$~mas) of the $15$-GHz VLBI jet of S5~1928+738 as a helical structure, with the following geometry, given in a system $\mathcal{K}_\mathrm{j}$ with the $z_\mathrm{j}$ coordinate along the jet symmetry axis:
\begin{flalign}
& x_\mathrm{j}(u)\!=\!\frac{b}{2\pi }u\cos u, \nonumber \\
& y_\mathrm{j}(u)\!=\!\frac{b}{2\pi }u\sin u, \nonumber \\
& z_\mathrm{j}(u)\!=\!\frac{a}{2\pi }u,
\label{coniceqs}
\end{flalign}
where $u$ is the parameter of the helical structure, with $a$ and $b\ll a$ being the axial and radial growth rates of the jet per turn, respectively. This shape appears in projection in the plane of the sky. An adapted coordinate system $\mathcal{K}$ has its $x$ and $y$ axes pointing towards north and east, respectively, with its $z$ axis along the line of sight (LOS). The symmetry axis of the helical jet has polar angle $\iota_0$ (the inclination angle measured from the LOS, $z$-axis) and azimuthal angle $\lambda_0$ (measured from the north, $x$-axis). For a jet pointing towards us, $\iota_0 \in [0,\pi/2]$.

The coordinates in $\mathcal{K}_\mathrm{j}$ are related to those in $\mathcal{K}$ as follows:

\begin{flalign}
\begin{pmatrix}
x \\ 
y \\ 
z \\
\end{pmatrix}
= R_z(\lambda_0) R_y(\iota_0)
\begin{pmatrix}
x_\mathrm{j} \\ 
y_\mathrm{j} \\ 
z_\mathrm{j} \\
\end{pmatrix},
\label{eq:kjkl}
\end{flalign}
where the rotation matrices are
\begin{flalign}
R_y(\iota{_0}) \!=\!
\begin{pmatrix}
\cos \iota{_0} & 0 & \sin \iota{_0} \\ 
0 & 1 & 0 \\ 
-\sin \iota{_0} & 0 & \cos \iota{_0}
\end{pmatrix}
\label{riota}
\end{flalign}
and
\begin{flalign}
R_z (\lambda{_0})\!=\!
\begin{pmatrix}
\cos \lambda{_0} & -\sin \lambda{_0} & 0 \\ 
\sin \lambda{_0} & \cos \lambda{_0} & 0 \\ 
0 & 0 & 1%
\end{pmatrix}.
\label{rlambda}
\end{flalign}
Hence, the coordinates of the helical structure in the plane of sky, employing equation (\ref{coniceqs}) are 
\begin{flalign}
\label{proj_geom1}
& {x}(u)\!=\!F(u)\cos\lambda{_0}-G(u)\sin(\lambda{_0}),\\
& {y}(u)\!=\!F(u)\sin\lambda{_0}+G(u)\cos(\lambda{_0}),
\label{proj_geom2}
\end{flalign}
where
\begin{flalign}
& F(u)\!=\!u \left(\frac{b}{2 \pi} \cos( u)\cos(\iota{_0})+\frac{a}{2\pi}\sin(\iota{_0}) \right),\\
& G(u)\!=\!\frac{b}{2\pi} u\sin(u).
\end{flalign}
Then the inclination angle and the position angle of the helical jet in the source frame arise as
\begin{align}
\iota(u)&=\arcsin{\sqrt{x(u)^2+y(u)^2}},\\
\lambda(u)&=\arctan \left({y(u)}/{x(u)}\right).
\end{align}

The intrinsic half-opening angle of the jet cone ($\psi_0$) can be given by the radial and axial increments as 
\begin{flalign}
\tan \psi_0\!=\!\frac{\sqrt{{x_\mathrm{j}}^2+{y_\mathrm{j}}^2}}{z_\mathrm{j}}=\frac{b}{a}.
\end{flalign}

The apparent half-opening angle of the cone $\psi^\mathrm{app}$ is obtained through the following reasoning. The generating lines of the cone 
\begin{flalign}
\hat{G}_{\mathrm{j}\pm}=
\begin{pmatrix}
\sin \psi_0 \cos (\lambda_0 \pm\frac{\pi}{2}) \\ 
\sin \psi_0 \sin (\lambda_0 \pm\frac{\pi}{2}) \\ 
\cos \psi_0 \\
\end{pmatrix},
\end{flalign}
which lie in the plane perpendicular to the ($x_\mathrm{j}$, $y_\mathrm{j}$) projection of the LOS are transformed into the system $\mathcal{K}$ system through $\hat{G}_\pm= R_y(\iota{_0}) \hat{G}_{\mathrm{j}\pm}$, obtaining

\begin{flalign}
\hat{G}^{x}_\pm\!&=\! \sin \iota_0 \cos \psi_0 +\cos \left(\lambda_0\pm \frac{\pi}{2}\right)\cos \iota_0  \sin \psi_0, \\
\hat{G}^{y}_\pm&= \sin \left(\lambda_0 \pm \frac{\pi}{2} \right) \sin \psi_0,  \\
\hat{G}^{z}_\pm\!&=\!\cos \iota_0 \cos \psi_0-\cos \left(\lambda_0\pm \frac{\pi}{2}\right)\sin \iota_0 \sin \psi_0.
\end{flalign}
Note that the scalar product
\begin{equation}
    \hat{G}_{\mathrm{j}+} \cdot \hat{G}_{\mathrm{j}-}=\cos 2\psi_0
\end{equation}
gives the intrinsic opening angle, while the projections of $\hat{G}_\pm$ to the ($x,y$) plane span the apparent opening angle $2\psi^\mathrm{app}$ according to
\begin{equation}
    \hat{G}^x_+ \hat{G}^x_- + \hat{G}^y_+ \hat{G}^y_-=\cos 2\psi^\mathrm{app}.
\end{equation}

The apparent half-opening angle of the jet cone $\psi^\mathrm{app}$ is related to its intrinsic value $\psi_0$ as $(0\degr<\iota _{0}<90\degr, \psi_0<\iota_0)$
\begin{flalign}
\tan \psi^\mathrm{app}\!=\!\frac{ \tan \psi_0}{\sin \iota_0} \left(1-\frac{\tan^2 \psi_0}{\tan^2 \iota_0}\right)^{-1/2}.
\end{flalign}
For small values of $\psi_0$ the above equation simplifies (in radians) as
\begin{equation}
    \psi ^\mathrm{obs} \approx \frac{\psi_0}{\sin \iota_0}.
    \label{psziobs}
\end{equation}

In \citet{Kun2014}, we derived the parameters $a=(10.6\pm 0.4)$~mas, $b=(0.18\pm0.06)$~mas and the initial values $\iota_0=7.0\degr$ and $\lambda_0=(160\pm 2)\degr$ (${\chi^2}_\mathrm{red}=0.31$, $D=11$). Here and throughout the paper, we used the Levenberg--Marquardt algorithm for fitting purposes, such that the $\chi^2$ was minimised during the fitting. Then keeping $a$ and $b$ as fixed parameters, we fitted the time variation of $\iota_0$ and $\lambda_0$ to the $15$-GHz VLBI jet components of the inner jet of S5~1928+738 in $43$ epochs between 1995.96 and 2013.06. The best-fit results showed that the variation of the inclination of the inner $2$~mas of the jet of S5~1928+738 is best described with a periodic term with an amplitude of $\sim 0.89\degr$ and a linear decreasing trend with a rate of $\sim 0.05\degr\,\mathrm{yr}^{-1}$. 

 \begin{table}
\begin{center}
\caption{The flux density of the 15 GHz jet ($F'$) and inclination of the jet axis ($\iota_0$) obtained from the parametric fit on the $15$-GHz data between 2013.34 and 2020.89. The full table is available in electronic format online.}
\begin{tabular}{@{}c c c}
\hline
\hline
Epoch & $F' (\mathrm{Jy})$ & $\iota_0 (\degr)$ \\
\hline
2013.34 & $3.405\pm 0.035$ & $9.81\pm1.04$\\
2013.61 & $3.626\pm 0.042$ & $9.50\pm 1.11$\\
2013.95 & $4.037\pm 0.029$ & $9.30\pm 1.03$\\
2014.66 & $4.878\pm 0.036$ & $7.65\pm 0.84$\\
2015.45 & $6.350\pm 0.037$ & $7.70\pm 0.85$\\
\hline
\hline
\end{tabular}
\label{table:fluxden_incli}
\end{center}
\end{table} 

We repeated the fit of the projected helical jet to the $(x,y)$ positions of the inner components of the $15$-GHz VLBI jet of S5~1928+738 employing new $15$-GHz VLBA observations in 29 additional epochs between 2013.34 and 2020.89. We present the best-fit results for $\iota_0$ in Table~\ref{table:fluxden_incli}. We show the phenomenological fit to the inclination angle change of the symmetry axis of the $15$-GHz inner jet of S5~1928+738 with a periodic and a monotonic term \citep[see in][]{Kun2014} described as
\begin{equation}
    \iota_0 (t)=A_0\!+\!A_1\sin \left(\frac{2\pi}{T}t\!-\!\phi \right)\!+\!A_2 t\\
\end{equation}
in Fig.~\ref{fig:phenofit}, where ${\chi^2}=37.8$, and the degrees of freedom $D=67$. We show the previous phenomenological fit that we got based on the inclination angle variation until 2013.06 (${\chi^2}=20.9$, $D=38$) and the prediction based on it, together with the extended time series and the fit on it. In Table \ref{fit_res}, we present the best-fit parameters of the phenomenological fit that we acquired in \citet{Kun2014} and the best-fit values of the same parameters that we got by fitting the extended time series of the inclination angle variation of the inner jet symmetry axis of S5~1928+738 (a total of 72 epochs between 1995.96 and 2020.89). The new fitted parameters agree within the errors of the previous parameters obtained in \citet{Kun2014}. Employing the extended inclination angle curve, we are able to constrain the errors on the fitted parameters more tightly. Most importantly, the relative error on the linear trend decreased from 45\% to 22\%, which makes it possible to constrain the velocity of the spin precession more precisely.

\section{The general jet emitter binary black hole model}
\label{sec:model}
\subsection{Newtonian order: the effect of orbital motion on the jet}
\label{subsec:0pn}
In our model, the jet velocity is the vectorial sum of the intrinsic jet velocity and the orbital velocity of the jet emitter black hole in the instant of the jet emission. In the following we define these vectors and derive the jet velocity.

\begin{figure*}
    \centering
    \includegraphics[width=0.75\textwidth]{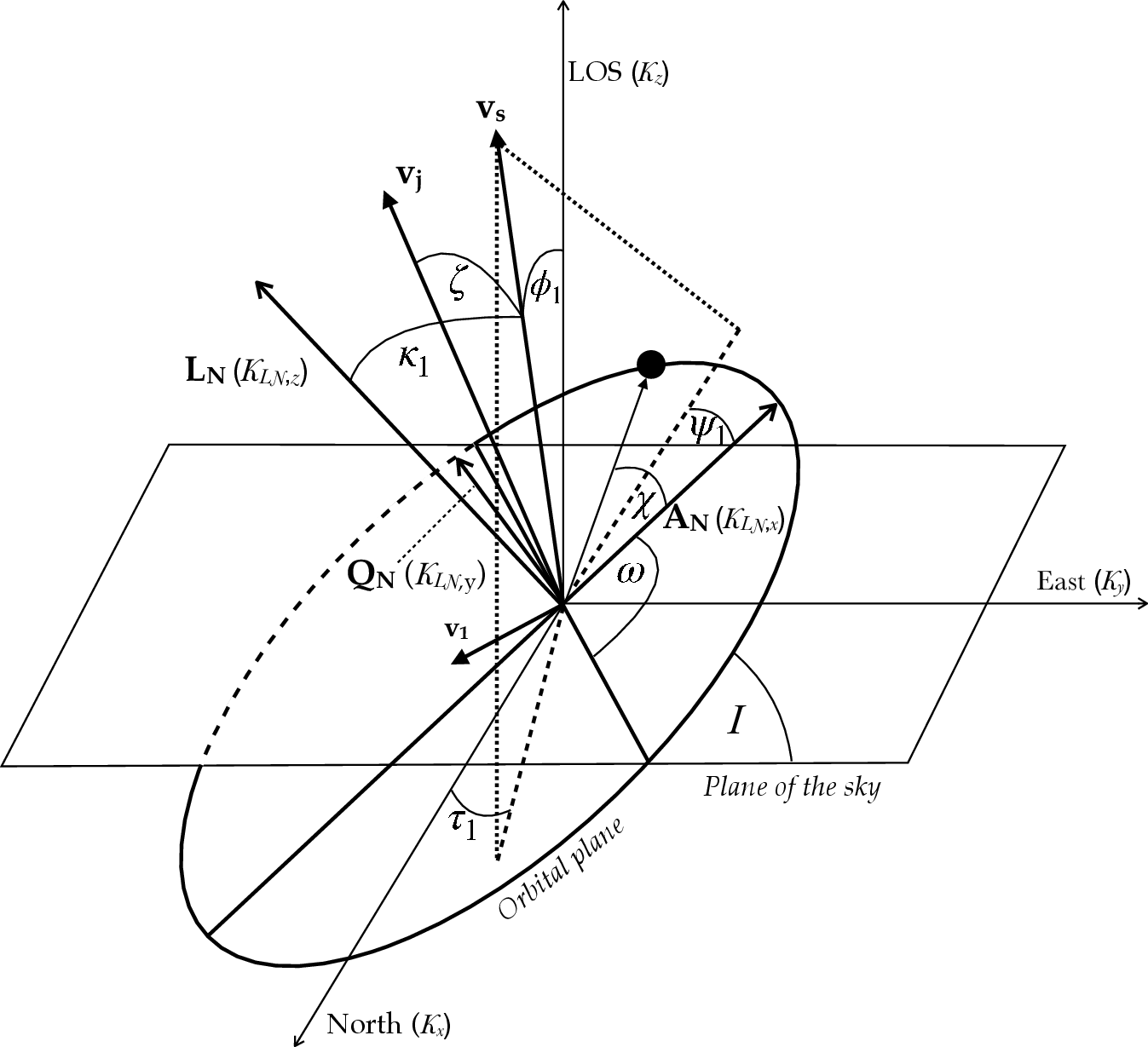}
    \caption{Geometric configuration of the black hole binary system centred on its barycentre. The black dot marks the position of the jet-emitting SMBH along its elliptical orbit. LOS indicates the line of sight, $\mathbf{L_\mathrm{N}}$ is the Newtonian orbital angular momentum, $\mathbf{A_\mathrm{N}}$ is the Laplace--Runge--Lenz vector, and $\mathbf{\hat{Q}_N}=\mathbf{\hat{L}_N}\times \mathbf{\hat{A}_N}$. The true anomaly $\chi$ is measured from $\mathbf{\hat{A}_N}$, the argument of the periapsis is $\omega$, the orbital inclination is $I$, the spin angle of the dominant SMBH (which is the jet emitter) with respect to the orbital normal is $\kappa_1$, the angle between the projection of the dominant spin onto the orbital plane and the periapsis line is $\psi_1$ and the inclination angle of the spin with respect to the LOS is $\phi_1$. The position angle of the spin projected onto the plane of the sky ($\tau_1$) is measured from north through east. Furthermore, $\mathbf{v_1}$ is the orbital velocity vector of the dominant SMBH at the instant of the jet component launching, $\mathbf{v_s}$ is the original jet velocity vector (that is parallel to the spin), and $\mathbf{v_\mathrm{j}}$ is the vectorial sum of these two. Finally, $\zeta$ is the instantaneous half-opening angle of the jet.}
    \label{fig:bbh_setup}
\end{figure*}

To describe geometry of our jet model, we define an orthonormal basis $\mathcal{K}_\mathrm{LN}$, such that the direction of the Laplace--Runge--Lenz vector $\mathbf{\hat{A}_N}$ and $\mathbf{\hat{Q}_N}=\mathbf{\hat{L}_N}\times \mathbf{\hat{A}_N}$ form the orbital plane together with $\mathbf{\hat{L}_N}$ (where $\mathbf{\hat{L}_N}$ is the direction of the Newtonian orbital angular momentum, perpendicular to the orbital plane). We show the geometric configuration of the black hole binary system fixed on its barycentre in Figure \ref{fig:bbh_setup}.

The radial vector of the reduced mass in the orbital plane is given as
\begin{gather}
\mathbf{r}
 = 
  \begin{pmatrix}
   r\cos\chi \\
   r\sin\chi\\
   0
   \end{pmatrix}.
   \label{eq:v1}
\end{gather}

Then orbital velocity of the reduced mass in a black hole binary is derived as:
\begin{gather}
\mathbf{v}
 = 
  \begin{pmatrix}
   \dot{r} \cos{\chi}-r \dot{\chi} \sin \chi \\
   \dot{r} \sin{\chi}+r \dot{\chi} \cos \chi\\
   0
   \end{pmatrix},
\end{gather}
with
\begin{align}
\dot{\chi}&=\frac{L_\mathrm{N}}{\mu r^2},\\
\dot{r}&=\frac{A_\mathrm{N}}{L_\mathrm{N}} \sin \chi,\\
r&=\frac{p}{1+e\cos\chi},\\
p&=\frac{L_\mathrm{N}^2}{Gm\mu^2},\\
e&=\frac{A_\mathrm{N}}{Gm \mu},\\
\end{align}
where $L_\mathrm{N}$ is the length of the Newtonian orbital momentum vector $\mathbf{L_\mathrm{N}}$, $A_\mathrm{N}$ is the length of the Laplace--Runge--Lenz vector $\mathbf{A}_\mathrm{N}$, $\chi$ is the true anomaly (measured from the direction $\mathbf{\hat{A}_\mathrm{N}}$ in the plane of motion), $e$ is the eccentricity, $G$ is the gravitational constant, $m=m_1+m_2$ ($m_1>m_2$) is the total mass, and $\mu=m_1 m_2/m$ is the reduced mass.

Then the velocity of the reduced mass in $\mathcal{K}_\mathrm{LN}$ becomes
\begin{gather}
\mathbf{v}
 = 
 \frac{G m \mu}{L_\mathrm{N}}
  \begin{pmatrix}
   -\sin \chi  \\
   e+\cos \chi\\
   0
   \end{pmatrix},
   \label{eq:redseb}
\end{gather}
where $E$ is the eccentric anomaly that is related to the true anomaly $\chi$ as
\begin{align}
\chi=2\arctan \left[\sqrt{\frac{1+e}{1-e}}\tan \frac{E}{2}\right].
\end{align}
The Kepler equation gives the time dependence of the eccentric anomaly as
\begin{align}
    E-e\sin E=n t,
\end{align}
with the mean anomaly $n=2\pi /T$. 

The two orbiting bodies form a gravitationally bound system, in which the bodies orbit under the forces in 0PN order (Newton's second law) 
\begin{flalign}
    m_1 \ddot{\mathbf{r_1}}&=\mathbf{F_{12}},\\
    m_2 \ddot{\mathbf{r_2}}&=\mathbf{F_{21}}.
\end{flalign}
Since $\mathbf{F_{12}}$ and $\mathbf{F_{21}}$ are equal in length and opposite in direction ($F_{12}+F_{21}=0$, $\mathbf{r_1}-\mathbf{r_2}=\mathbf{r}$), one can derive the position of the orbiting bodies as
\begin{flalign}
    \mathbf{r_1}&=\frac{m_2}{m} \mathbf{r},\\
    \mathbf{r_2}&=-\frac{m_1}{m} \mathbf{r},
\end{flalign}
where $\mathbf{r_1}$ and $\mathbf{r_2}$ point to the locations of $m_1$ and $m_2$ from the position of the centre of mass that travels with constant velocity ($\mathbf{r_1}=\mathbf{R_1}-\mathbf{R_{CM}}$ and $\mathbf{r_2}=\mathbf{R_2}-\mathbf{R_{CM}}$, where $\mathbf{R_{CM}}$ is the location of the centre of mass, while $\mathbf{R_1}$ and $\mathbf{R_2}$ are vectors describing the motion of $m_1$ and $m_2$).
From this, the velocity of the two bodies are
\begin{gather}
\mathbf{v_1}
 = \frac{m_2}{m}
 \mathbf{v},
\end{gather}
\begin{gather}
\mathbf{v_2}
 = -\frac{m_1}{m} \mathbf{v},
\end{gather}
where $\mathbf{v}$ is the relative velocity between $\mathbf{v_1}$ and $\mathbf{v_2}$. Then, from equation (\ref{eq:redseb})

\begin{gather}
\mathbf{v_1}
 = 
  \frac{G m \mu}{L_\mathrm{N}}\frac{\nu}{(1+\nu)}
  \begin{pmatrix}
   -\sin \chi  \\
   e+\cos \chi\\
   0
   \end{pmatrix},
\end{gather}
\begin{gather}
\mathbf{v_2}
 = -  \frac{G m \mu}{L_\mathrm{N}}\frac{1}{(1+\nu)}
  \begin{pmatrix}
   -\sin \chi  \\
   e+\cos \chi\\
   0
   \end{pmatrix},
\end{gather}

As mentioned above, the jet velocity vector $\mathbf{v_{j}}$ in $\mathcal{K}_\mathrm{LN}$ is the vector sum of $\mathbf{v_{s}}$ and the orbital velocity of the jet emitter black hole $\mathbf{v_1}$ (both in $\mathcal{K}_\mathrm{LN}$), i.e. $\mathbf{v_j}=\mathbf{v_s}+\mathbf{v_1}$. Now we derive $\mathbf{v_j}$.

The jet power ($P_\mathrm{j}$) is predicted to depend on the spin of the black hole as $P_\mathrm{j}\approx (SB)^2$ \citep{BlandfordZnajek1977}, where $S=aM$ is the spin, $a$ is the dimensionless spin parameter, $M$ is the mass), and $B$ is the magnetic field strength at the horizon of the black hole. Indeed, observations of radio-loud AGN show that the accretion disk alone is not enough to power up the relativistic jets of AGN \citep{Ghisellini2014}. It is plausible to assume SMBHs in a tight binary system are co-evolving in a similar environment acquiring similar dimensionless spin parameters. In case of similar rotation velocities the ratio of the black hole spins, therefore the ratio of the jet power in a double-jetted binary system depends mostly on the mass ratio of the black holes. Since we do not see the signs of another jet in S5~1928+738 that would indicate the presence of a comparable powered pair of jets, we assume the dominant SMBH is the jet emitter black hole, similarly to our earlier work \citep{Kun2014}. In the $\mathcal{K}$ coordinate system (attached to the LOS, see in Section \ref{sec:jetorientation}), the inclination and position angle of the dominant spin $\mathbf{S_1}$ are $\phi_1$ and $\tau_1$ (see Fig. \ref{fig:bbh_setup}). In 0PN order, the direction of the dominant spin does not change with time $\dot\phi_1^{0PN}=\dot\tau_1^{0PN}=0$. In $\mathcal{K}_\mathrm{LN}$, the spin of the jet emitter BH $\mathbf{S_1}$ points to the direction
\begin{gather}
\mathbf{S_\mathrm{1}}=
\begin{pmatrix}
  v_\mathrm{s} \sin \kappa_1 \cos \psi_1\\ 
 v_\mathrm{s} \sin \kappa_1 \sin \psi_1\\ 
 v_\mathrm{s} \cos \kappa_1
\end{pmatrix},
\end{gather}
with the polar angle $\kappa_1= \arccos (\mathbf{\hat{S}_1} \cdot \mathbf{\hat{L}_N})$ measured between $\mathbf{S_1}$ and $\mathbf{L_N}$ and azimuthal angle $\psi_1$ measured between the projection of the dominant spin $\mathbf{{S}_1}$ onto the orbital plane and the periapsis line. Assuming that the dominant SMBH emits the jet via the Blandford--Znajek mechanism \citep{BlandfordZnajek1977}, the unperturbed jet velocity is then $\mathbf{v_\mathrm{s}}=v_\mathrm{s} \mathbf{\hat{S}_1}$. 

The jet velocity vector $\mathbf{v_{j}}$ in $\mathcal{K}_\mathrm{LN}$ is the vector sum of $\mathbf{v_{s}}$ and the orbital velocity $\mathbf{v_1}$ (both in $\mathcal{K}_\mathrm{LN}$), such that
\begin{gather}
\mathbf{v_\mathrm{j}}(\mathcal{K}_\mathrm{LN})=
\begin{pmatrix}
 v_\mathrm{s} \sin \kappa_1 \cos \psi_1-v_{1} \sin \chi\\ 
 v_\mathrm{s} \sin \kappa_1 \sin \psi_1+v_{1} (e+\cos  \chi)\\ 
 v_\mathrm{s} \cos \kappa_1
\end{pmatrix},
\end{gather}
where
\begin{equation}
    v_1=\frac{G m \mu}{L_\mathrm{N}}\left(\frac{\nu}{1+\nu}\right).
\end{equation}
Then the half-opening angle of the precession cone induced by the orbital motion in the source frame $\zeta$ can be given as
\begin{align}
    &\cos \zeta= \left[ \frac{\mathbf{v_{j}^2+\mathbf{v_\mathrm{s}}^2}-\mathbf{v_1}^2}{2 {v_\mathrm{s}} {v_\mathrm{j}}} \right]=\nonumber \\
    \label{eq:zeta}
    &\frac{1+X  f(\kappa,e,\chi,\psi_1)}
    {((1+e^2)X^2+1+2 e X^2 \cos \chi+2X f(\kappa,e,\chi,\psi_1))^{1/2}},
\end{align}
with
\begin{equation}
    X=\frac{v_1}{v_\mathrm{s}}
\end{equation}
and
\begin{equation}
    f(\kappa_1, e,\chi,\psi_1)=\sin\kappa_1 \left[e \sin\psi_1-\sin{(\chi-\psi_1)}\right].
    \label{eq:fterm}
\end{equation}

Since $\zeta$ and $\chi$ correspond to the polar and azimuthal angles of the vector $\mathbf{v_j}$, respectively, in a spherical polar system, we can write down the jet velocity given in $\mathcal{K}_\mathrm{j}$ as
\begin{gather}
\mathbf{v_\mathrm{j}}(\mathcal{K}_\mathrm{j})=v_\mathrm{j}
\begin{pmatrix}
 -\sin \zeta\sin\chi\\ 
 \sin \zeta(e+\cos \chi)\\ 
 \cos \zeta
\end{pmatrix},
\end{gather}
where $\zeta$ is the instantaneous half-angle of the precession cone and $\chi$ is the instantaneous true anomaly. 

The jet velocity in $\mathcal{K}_\mathrm{j}$ is related to that in $\mathcal{K}$ as follows:
\begin{flalign}
\begin{pmatrix}
v_{\mathrm{j},x}'' \\ 
v_{\mathrm{j},y}'' \\ 
v_{\mathrm{j},z}'' \\
\end{pmatrix}
= R_z(\tau_1) R_y(\phi_1)
\begin{pmatrix}
v_{\mathrm{j},x}  \\ 
v_{\mathrm{j},y}  \\ 
v_{\mathrm{j},z} \\
\end{pmatrix},
\end{flalign}
where the forms of $R_y(\phi_1)$ and $R_z(\tau_1)$ are given by equation (\ref{riota}) and equation (\ref{rlambda}), respectively. Then the jet velocity projected in the plane of the sky becomes 
\begin{flalign}
v_{\mathrm{j},x}''={v_{\mathrm{j},x}'}\cos \tau_1^\mathrm{0PN}- {v_{\mathrm{j},y}'}\sin \tau_1^\mathrm{0PN}, \\ 
v_{\mathrm{j},y}''={v_{\mathrm{j},x}'}\sin \tau_1^\mathrm{0PN}+ {v_{\mathrm{j},y}'}\cos \tau_1^\mathrm{0PN},
\end{flalign}
where
\begin{flalign}
v_{\mathrm{j},x}'&=-v_\mathrm{j}\sin \zeta\sin\chi\cos \phi_1^\mathrm{0PN}+v_\mathrm{j}\cos \zeta\sin \phi_1^\mathrm{0PN}, \\ 
v_{\mathrm{j},y}'&=v_\mathrm{j} \sin \zeta (e+\cos\chi).
\end{flalign}

Then the inclination angle and the position angle of the jet ejection in 0PN order:
\begin{align}
\label{eq:iota}
\iota_\mathrm{j}^\mathrm{0PN}&=\arcsin{\sqrt{{v_\mathrm{jx}''}^2+{v_\mathrm{jy}''}^2}},\\
\lambda_\mathrm{j}^\mathrm{0PN}&=\arctan ({v_{\mathrm{j},y}''}/{v_{\mathrm{j},x}''}).
\end{align}

Making a series expansion in $\zeta$, keeping only the leading order (since the orbital velocity of the jet emitter black hole is much smaller then the jet velocity, and $\zeta$ should be small), we get the inclination angle of the jet ejection from equation (\ref{eq:iota}) (for $\phi_1^{0PN}>0$) in the source frame as:
\begin{align}
    &\sin \iota_\mathrm{j}^{0PN}=\sin \phi_1^\mathrm{0PN}-\zeta\cos \phi_1^\mathrm{0PN} \sin\chi,
    \label{eq:inull0pnproj}
\end{align}
where $\zeta$ comes from equation (\ref{eq:zeta}). 

\subsection{The 1.5~PN order: inclusion of the spin precession}
\begin{figure*}
    \centering
    \includegraphics[width=0.6\textwidth,angle=270]{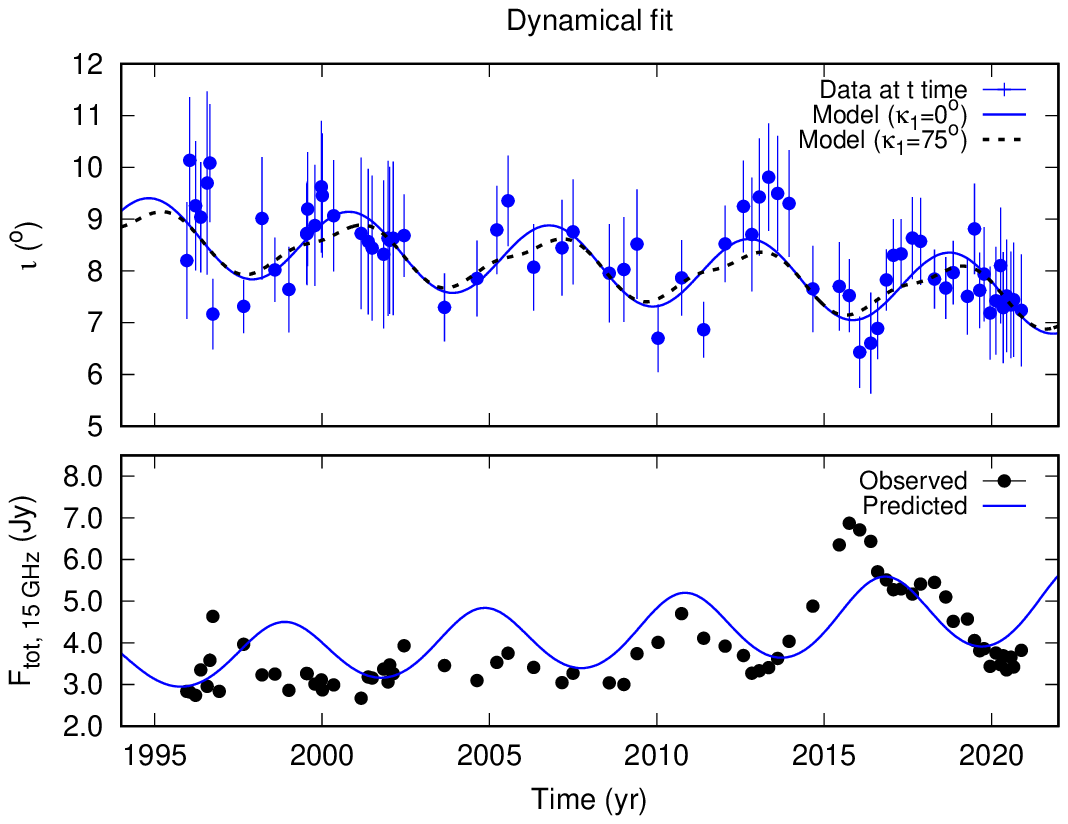}
    \caption{Upper panel: Inclination variation of the symmetry axis of the $15$-GHz jet of S5~1928+738 (blue dots with error bars), and the dynamical fit with $\kappa_1=0\degr$ (blue continuous line) and with $\kappa_1=75\degr$ (black dashed line). Lower panel: Observed total flux density of the $15$-GHz VLBI jet of S5~1928+738 (black dots with errors, note the error bars are usually smaller than the symbols), overlaid with the prediction based on the dynamical fit of the inclination variation of the symmetry axis of the inner jet (blue continuous line).}
    \label{fig:incl_change}
\end{figure*}
Up to the second PN order, the binary dynamics is conservative, the total energy $E$ and the total momentum $\mathbf{J}=\mathbf{S}_1+\mathbf{S}_2+\mathbf{L_N}$ being the constants of motion. If the $\mathbf{S}_1$ and $\mathbf{S}_2$ spins ($m_1>m_2$) are not parallel with the Newtonian orbital momentum $\mathbf{L_N}$, the spins begin to precess \citep{Barker1975,Barker1979}
\begin{align}
\mathbf{\dot{S}}_i\!=\!\Omega_i \times \mathbf{S}_i,
\end{align}
where $\mathbf{\Omega_{i}}$ is the angular momentum of the \textit{i} spin and it contains spin-orbit (1.5PN), spin-spin (2PN) and quadrupole-monopole contributions (2PN) up to 2PN order. In higher PN orders magnetic dipole, self-spin and back-radiation effects appear in the binary dynamics. Due to the gravitational radiation, the eccentricity of the orbits disappears \citep{Peters1964}, therefore from here we take only circular orbits into account.

\citet{Gergely2009} have shown that the typical mass ratio range of the merging supermassive black holes is $\nu \in [1/30:1/3]$, where the spin-flip of the dominant spin happens sometime during the binary evolution. 
In the one-spin case, when the second spin $S_2$ is neglected due to $\nu \in [1/30:1/3]$, the Newtonian orbital momentum $\mathbf{L_{N}}=\mu \mathbf{r}\times \mathbf{v}$ (where $\mathbf{v}$ is the velocity of the reduced mass $\mu$), the dominant spin $\mathbf{S}_1$ and the total momentum $\mathbf{J}$ lay in the same plane, such that
\begin{align}
 \alpha_1&=\arccos (\mathbf{\hat{L}_N} \cdot \mathbf{\hat{J}}),\\
\beta_1&=\arccos (\mathbf{\hat{S}_1} \cdot \mathbf{\hat{J}}),\\ 
\kappa_1&=\arccos (\mathbf{\hat{S}}_{1}\cdot \mathbf{\hat{L}_N}),   
\end{align}
where $\kappa_1=\alpha_1+\beta_1$ is the spin angle. Then the angular velocity vector $\mathbf{\Omega_1}$ of the dominant spin $\mathbf{S_1}$ with only spin-orbit (SO) contribution in 1.5PN order \citep{Gergely2010a} is:
\begin{align}
\label{spin_prec_1}
\mathbf{\Omega_1}^\mathrm{SO}&=\frac{G(4+3\nu)}{2c^2 r^3}L_\mathrm{N}\mathbf{\hat{L}_N},
\end{align}
where $r=|\mathbf{r}|$ is the separation along the unit vector $\mathbf{\hat{r}}=\mathbf{r}/|\mathbf{r}|$ and $e=0$, from which  
\begin{align}
\Omega_1^\mathrm{SO}&=\frac{c^{3}}{2Gm}\varepsilon ^{5/2}\eta \left(
4+3\nu \right).
\end{align}
\citet{Gergely2009} have shown that in the inspiral phase of the merger, when $L_\mathrm{N}>S_1$ holds, the radiation time-scale of the orbital momentum due to the gravitational waves and the angular velocity of the spin-orbit precession are:
\begin{align}
\frac{1}{T_\mathrm{GW}} &=-\frac{\dot{L}}{L}=\frac{32c^{3}}{5Gm}\varepsilon ^{4}\eta,\\
\Omega_\mathrm{p} &=\frac{2c^{3}}{Gm}\varepsilon ^{5/2}\eta,
\label{eq:timescales_so}
\end{align}%
respectively, where $\eta=\nu (1+\nu )^{-2}\in [0:0.25]$ is the asymmetrical mass ratio. 

In Section \ref{subsec:0pn}, we derived the inclination under which the jet ejects the blobs $\iota_\mathrm{j}^\mathrm{0PN}$ ($\dot{\iota}_\mathrm{j}^\mathrm{0PN}=0$), such that a periodicity in this direction arose due to the orbital motion. This angle, described by equation (\ref{eq:iota}), is valid only in the 0PN order while the dominant spin does not move and $\dot \phi_1^{0PN}=0$. This changes in 1.5 PN order because the spin precesses due to the spin-orbit precession and $\dot{\phi_1}^\mathrm{1.5PN}\neq 0$. Considering that the angle between the spin and the total angular momentum $\beta_1= \arccos (\mathbf{\hat{S}_1} \cdot \mathbf{\hat{J}})$ does not change in the conservative dynamics ($\dot{\beta}_1=0$) and that the jet emission alone \citep{BlandfordZnajek1977} can carry away the spin only typically over the Hubble time \citep{Armitage1999,Mangalam2009}, we can approximate the inclination angle of the dominant spin such that (similarly to equation \ref{eq:inull0pnproj})
\begin{equation}
    \phi_1^\mathrm{1.5PN}(t)=j_0+\beta_1 \cos j_0\sin \Omega_\mathrm{p} t,
    \label{eq:inull1.5pnproj}
\end{equation}
where $j_0$ is the inclination of $\mathbf{J}$. Inserting equation (\ref{eq:inull1.5pnproj}) to equation (\ref{eq:inull0pnproj}), the inclination angle of the jet symmetry axis (or the direction of the jet ejection in case of an intrinsically straight jet ridge line) becomes
\begin{align}
    \sin \iota_\mathrm{j}^\mathrm{1.5PN}(t)&=\sin \phi_1^\mathrm{1.5PN}(t)-\zeta(t)\cos \phi_1^\mathrm{1.5PN}(t) \sin(\chi-\psi_1),\nonumber \\
    \label{eq:final_incl}
   \iota_\mathrm{j}^\mathrm{1.5PN}(t) &\approx j_0+\beta_1 \cos j_0\sin \Omega_\mathrm{p} t - \zeta(t) \sin(\chi-\psi_1),
\end{align}
in radians, where we used the first order paraxial approximation of $\phi_\mathrm{1.5PN}(t)$ since $\iota_0<10^\circ$ all the time, see Fig. \ref{fig:phenofit}. For circular orbits, the true anomaly equals the mean anomaly $\chi=n t$, where $n=2\pi/T$ with $T$ being the orbital period. Therefore for circular orbit equation (\ref{eq:zeta}) simplifies to
\begin{equation}
  \cos\zeta(t)=\frac{1-X f_\mathrm{c}}{(X^2-2X f_\mathrm{c})^{1/2}},
\end{equation}
where $f_\mathrm{c}=\sin \kappa_1 \sin (\chi-\psi_1)$ (setting $e=0$ in equation \ref{eq:fterm})
and
\begin{align}
v_{1}=\sqrt{\frac{Gm}{r}}\left(\frac{\nu}{1+\nu}\right)
\end{align}
is the circular velocity of the jet emitter (dominant) black hole. Then, after correcting for the cosmological redshift, equation (\ref{eq:final_incl}) can be fitted to the inclination angle of the symmetry axis of the (intrinsically) helical jet. 

\begin{figure}
    \centering
    \includegraphics[angle=270,width=0.5\textwidth]{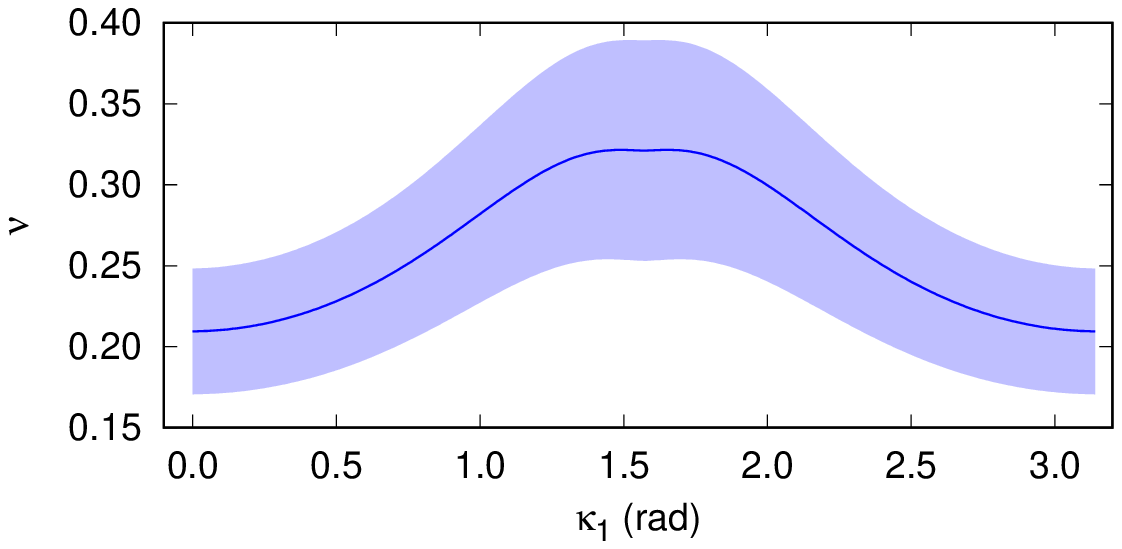}
    \caption{The best-fit mass ratio $\nu$ and its $\pm 1\sigma$ error for a given spin angle $\kappa_1$.}
    \label{fig:kappa_nu}
\end{figure}

\section{Test and results}
\label{sec:test}
\subsection{When the dominant spin is perpendicular to the orbital plane ($\kappa_1=0$)}
\label{subsec:kappa0}
In Section \ref{morphkin}, we derived the observed period as $T'=(5.98\pm0.10$) yr, from which, considering the redshift of S5~1928+738 as $z_\mathrm{spect}=0.302$, we get the orbital period in the source frame as $T\approx4.6$~yr. Employing the total mass $m=8.13\times 10^8 M_\odot$ as in \citet{Kun2014}, we get the orbital separation in the source frame as $r\approx0.013$~pc and then the PN parameter becomes $\varepsilon\approx0.003$. From jet kinematics, we identify $v_\mathrm{s}=0.992$. We define
\begin{equation}
\label{eq:vovos}
    \frac{v_1}{v_\mathrm{s}}=\frac{\varepsilon^{1/2}}{\beta_\mathrm{s}}\left(\frac{\nu}{1+\nu}\right)=K\left(\frac{\nu}{1+\nu}\right)=0.055 \left(\frac{\nu}{1+\nu}\right),
\end{equation}
and then equation (\ref{eq:final_incl}) becomes
\begin{align}
\label{eq:fittedincl}
    &\iota_\mathrm{j}^\mathrm{1.5PN}(t)= j_0+\beta_1 \sin \Omega_\mathrm{p} t-(1+z_\mathrm{spect})\nonumber \\
    &\times\arccos\left[\frac{1-K \left(\frac{\nu}{1+\nu}\right) \sin \kappa \sin (n't-\psi_1)}{\left(K^2 \left(\frac{\nu}{1+\nu}\right)^2+1- 2 K \left(\frac{\nu}{1+\nu}\right)\sin \kappa \sin (n't-\psi_1)\right)^{1/2}}\right]\nonumber \\
    &\times \sin(n't-\psi_1).
\end{align}
From our fit in Section \ref{morphkin} (see Table \ref{fit_res}), we identified the constant term as ${A_0}={j_0}=(8.64\pm0.16)\degr$, the linear trend as $A_2=\beta_1 \cos j_0 \sin \Omega_\mathrm{p}=(-0.044\pm0.01)\degr$\,yr$^{-1}$, the observed period as $T'=(5.98\pm0.10)$~yr, and the spin angle as $\phi=\psi_1= (24.9\pm 16.1)\degr$. Then we fitted equation (\ref{eq:fittedincl}) to the variation of the inclination angle of the symmetry axis of the $15$-GHz VLBI jet of S5~1928+738 (see in Fig.~\ref{fig:incl_change}.), keeping $\kappa_1$ zero (i.e. the dominant spin is perpendicular to the orbital plane, $\mathbf{S_1}|| \mathbf{L_N}$) and $K=0.055$ (see equation (\ref{eq:vovos})). Then the best-fit mass ratio emerged as $\nu (\kappa=0)=0.209 \pm 0.041$.

\subsection{When the dominant spin direction is arbitrary ($0 \leq \kappa_1 \leq \pi$): constrains on the mass ratio}

We repeated the test described in Section \ref{subsec:kappa0}, but now fitting the mass ratio for fixed values of $\kappa_1$ between $0$ and $\pi$. We show the resulted $\nu$ ($\kappa_1$) values and their $\pm 1\sigma$ errors in Fig. \ref{fig:kappa_nu}. The reduced $\chi^2$ of the fits changes between $\chi^2_\mathrm{red}=0.53$ and $\chi^2_\mathrm{red}=0.55$, indicating that we cannot choose a best model solely based on the $\chi^2$ values. We can constrain the mass ratio from $\nu=0.21\pm0.04$ ($\kappa_1=0$) to $\nu=0.32\pm0.07$ ($\kappa_1=\pi/2$).

\subsection{Sanity check with the total flux density}
We show the total flux density of S5~1928+738 observed at 15 GHz in the lower panel of  Figure \ref{fig:incl_change}. This radiation comes dominantly from the inner jet.
The apparent flux density $F'$ depends on the Doppler factor $\delta=\gamma^{-1} (1-\beta \cos \iota)^{-1}$, such that 
\begin{equation}
F'(\nu)=F (\nu)\delta^{n-\alpha} ,    
\end{equation}
where $F\sim \nu^{\alpha}$ is the intrinsic flux density, $\alpha$ is the spectral index and the value of $n$ in the exponent depends on the shape of the jet structure \citep[$n=2$ for a linear jet and $n=3$ for a spherical outflow, see in, e.g.][]{Urry1995}. Since S5~1928+738 has a long jet, in our model $n=2$. We estimated the apparent flux density variation of the inner jet of S5~1928+738 by substituting the fitted inclination angle variation of the jet symmetry axis $\iota(t)$ into the Doppler factor as $\delta(t)=\gamma^{-1} (1-\beta \cos \iota(t))^{-1}$. We already derived the characteristic jet velocity as $\beta\approx0.992$ (from which $\gamma=(1-\beta^2)^{-1/2}\approx7.92$ ) and a flat spectrum ($\alpha\approx -0.054$ for S5~1928+738, see in \citet{Kun2014}). Since we have no information on the intrinsic flux-density, this is the only free parameter of the fit. We show the resulted $F'(t)$ curve in the lower panel of Fig. \ref{fig:incl_change}, where the best-fit flux density in the source frame is $F=0.083\pm0.002$~Jy. Though the fit is only marginal in terms of the statistical goodness (${\chi^2}_\mathrm{red}\approx 219$), we still got back the main features of the observed flux density variations solely by assuming the best-fit parameters from our inclination fit $\iota(t)$ and freeing the intrinsic flux density.

There might be several reasons to explain why the prediction of the flux density variation based on the binary model is somewhat poor. Once, our jet model does not contain synchrotron emissivity structure (e.g. a non-uniformly distributed electron population, for example a spine--sheath structure), it is a kinematic model of the time variation of the symmetry axis of the inner jet.  In our model, the jet components are ejected either under smaller or larger viewing angles compared to the symmetry axis because of the inner helicity. Therefore deviations from the average behaviour, ups and downs on the average flux density curve, are expected.

\section{Discussion and conclusions}
\label{discsum}

In \citet{Kun2014}, we argued that the periodic and monotonic terms in the change of the inclination of the 15 GHz inner jet symmetry axis of S5~1928+738 can be best explained by spin-orbit precession. Jet precession arising from other phenomena, e.g. Lense--Thirring (LT) precession, magnetic instabilities, water-hose models, etc. could explain only either the periodic, or the monotonic term, but not both simultaneously.

The jet of a rotating black hole can precess due to the LT precession of a massive accretion disk that leads to the Bardeen--Petterson effect \citep{Bardeen1975}. In this case, the viscosity of a massive accretion disk of a Kerr black hole, combined with the LT effect, forces the innermost region of the accretion disk to be parallel to the rotation axis of the black hole \citep[see e.g.][]{Liska2019}, while the outermost regions of the accretion disk keep the original direction of the momenta. This combined effect leads to the precession of the accretion disk and to the precession of the jet \citep{BlandfordZnajek1977,Liska2018}. In this case, the direction of the symmetry axis of the jet helix changes either periodically, or in a monotonic way, depending on the time-scale of the precession \citep[see an application of the LT model in, e.g., Figure 3 of][]{Caproni2006}. This is contrary to our results, because what wee see is a periodic and monotonic change, acting together on the jet structure.

Magnetohydrodynamic (MHD) instabilities, most prominently Kelvin--Helmholtz (KH) instabilities, can also lead to wiggled jets \citep[for application, see e.g.][]{Perucho2006}. In this case, what we see is not a pattern motion, but jet components moving along curved paths. This might be the cause of the internal helicity of the jet in S5 1928+738. What we see in the case of S5 1928+738 is that the direction of the symmetry axis of this jet helix changes periodically, on top of a much smaller, seemingly monotonic trend. Models relying on MHD (or specifically KH) instabilities, can not explain the observed complexity in S5 1928+738: A periodically changing inclination angle of the symmetry axis on top of a monotonic decrease.

Assuming $m=8.13\times 10^8 M_\odot$ as the total mass of the binary ($\varepsilon\approx0.003$, $r\approx0.0129$), we derived the orbital period in the source frame as $4.78$~yr and the lower limit on the mass ratio as $\nu_\mathrm{lo}=0.21$ from the jet kinematics and the periodic term of the inclination variation of the symmetry axis of the inner jet of S5 1938+738. For $\nu>0.21$, the spin-orbit period is $T_\mathrm{SO} <5500$~yr and the merger time is $T_\mathrm{GW} <1.64 \times 10^6$~yr. Assuming the upper limit of typical mass ratio range of merging SMBHs $\nu<1/3$, we constrained the spin-orbit period with lower limits as $T_\mathrm{SO} >4207$~yr and the merger time as $T_\mathrm{GW} >1.25 \times 10^6$~yr. 

We summarise our results regarding the $1.5$~PN dynamics of the hypothetical binary of S5 1928+738 in Table~\ref{tab:binarypars}. In the recent analysis, we derived the orbital period as $\approx4.6$~yr. Then, assuming the total mass as $m=8.13\times 10^8 M_\odot$, the orbital separation (for circular orbit) arises as $r\approx 0.0125$ pc. From the jet kinematics, we got $\nu=0.21\pm0.04$ for $\kappa_1=0$ (i.e. when the spin direction is perpendicular to the orbital plane) and $\nu=0.32\pm0.07$ for $\kappa_1=\pi/2$ (i.e. when the spin lies in the orbital plane). For the mass ratio $\nu=0.21\pm0.04$, the spin-orbit precession is $T_\mathrm{SO}=5171^{+802}_{-536}$~yr and the merger time is $T_\mathrm{GW}=(1.48^{+0.23}_{-0.15})\times 10^6$~yr. For the mass ratio $\nu=0.32\pm0.07$, the spin-orbit precession is $T_\mathrm{SO}=4039^{+597}_{-364}$~yr and the merger time is $T_\mathrm{GW}=(1.16^{+0.17}_{-0.10})\times 10^6$~yr. 

\begin{table}
    \centering
    \begin{tabular}{lccc}
       \hline
       \hline
        & $\nu$ & $T_\mathrm{SO}$ & $T_\mathrm{GW}$\\
        \hline
        $\kappa_1=0$ & $0.21\pm0.04$ & $5171^{+802}_{-536}$~yr & $(1.48^{+0.23}_{-0.15})\times 10^6$~yr\\
$\kappa_1=\pi/2$ & $0.32\pm0.07$ & $4039^{+597}_{-364}$~yr & $(1.16^{+0.17}_{-0.10})\times 10^6$~yr\\
        \hline
    \end{tabular}
    \caption{Estimation of the mass ratio, the spin-orbit precession period and the merger time for S5~1928+738, assuming the total mass as $m=8.13\times 10^8 M_\odot$ and the orbital period as $T\approx4.6$~yr in the source frame.}
    \label{tab:binarypars}
\end{table}

The dynamical evolution of a SMBH binary rotates the spin to be parallel with the Newtonian orbital momentum $\mathbf{L_N}$, which means the spin is perpendicular to the orbital plane. For $\kappa_1=0\degr$, it is an aligned, for $\kappa_1=\pi$, it is an anti-aligned configuration. Once $\kappa_1$ becomes $0$ or $\pi$, the configuration itself becomes stable. In this case, the spin-orbit precession disappears. Since we see the slow, secular reorientation of the jet symmetry axis (on top of a much faster periodicity, due to the orbital motion), our model is consistent with the observations and phenomenological model when we exclude the possibility of $\kappa_1=0$ and $\kappa_1=\pi$. 

Based on the periodic modulation and the monotonic trend in the inclination angle of the symmetry axis of the inner jet, we further supported the existence of the SMBH binary in the heart of the quasar S5~1928+738. We made a more precise estimation of the velocity of the spin precession, from $(-0.05\pm0.02)\degr$\,yr$^{-1}$ to $(-0.04\pm0.01)\degr$\,yr$^{-1}$. Using this rate, we estimate the symmetry axis of the jet goes through our line of sight sometime between 2180 and 2200, which is far from the present. Finally we conclude that the precession of the $15$-GHz VLBI inner jet of S5~1928+738 is still ongoing, showing a similar behaviour over $25$ years between 1995.96 and 2020.89. Since the jet is approaching our line of sight and the Doppler boosting strengthens non-linearly with the decreasing inclination angle, we advise to monitor the quasar S5~1928+738 across all regimes of the electromagnetic spectrum. Depending on the geometry, either the jet will go through our line of sight, or turn back toward larger inclination angles. The merging SMBBH system in S5 1928+738 is in the beginning of its inspiral phase, emitting gravitational waves at a frequency of $10^{-8}$ Hz and with a characteristic strain in the order of $10^{-14}$ \citep[based on the calculations in][]{Sesana2016}. Further multimessenger observations might prove further or challenge the existence of the SMBH binary harboured in S5 1928+738.

\section*{Acknowledgements}
We thank Zolt\'an Haiman for valuable comments on our earlier related work. E.K. thanks the Alexander von Humboldt Foundation for its Fellowship. We acknowledge support from the Deutsche Forschungsgemeinschaft DFG, within the Collaborative Research Center SFB1491 "Cosmic Interacting Matters - From Source to Signal" (project No. 445052434). Support by the Hungarian National Research, Development and Innovation Office (NKFIH) is acknowledged (grant numbers OTKA K134213 and K123996). This research has made use of data from the MOJAVE database that is maintained by the MOJAVE team \citep{Lister2018}. The National Radio Astronomy Observatory is a facility of the National Science Foundation operated under cooperative agreement by Associated Universities, Inc. 

\section*{Data Availability}
The data underlying this article are available at \url{https://www.cv.nrao.edu/MOJAVE/sourcepages/1928+738.shtml} as part of the MOJAVE Survey.

\end{document}